# High performance beam transport with multi-stage acceleration system and its application to plasma potential measurement in fusion plasmas

K. Nakamura[a,b], M. Nishiura[a,c], K. Ueda[c], A. Shimizu[c], H. Takubo[c], M. Kanda[c], T. Ido[d]

[a] Graduate School of Frontier Sciences, The University of Tokyo, Kashiwa, Chiba 277-8561, Japan
[b] Nishina Center for Accelerator-Based Science, RIKEN, Wako, Saitama 351-0198, Japan
[c] National Institute for Fusion Science, 322-6 Oroshi-Cho, Toki 509-5292, Japan
[d] Department of Advanced Energy Engineering, Kyushu University, Kasuga, Fukuoka 816-8580, Japan

**ABSTRACT**

In accelerators, ion beams are often accelerated using electrostatic accelerating tubes. This paper reports on a method to improve the beam transport efficiency without adding new components to the beam transport system. High beam currents often suffer from a beam loss in a transport line. When Au negative ion beams are injected into a tandem accelerator, numerical simulations of low-energy ion beam transport have found that the beam loss increases significantly when the Au or Cu negative ion beam current exceeds 100 μA due to space-charge effects. We found that the transport efficiency is significantly improved by remaining constant beam energy accelerated at the multi-stage accelerator tube and by providing an electrostatic lens effect. In the heavy ion beam probe system (HIBP) system of the Large Helical Device (LHD) for plasma potential measurement, the negative ion beam current injected into the tandem accelerator could be increased by a factor of 3.6. As a result, the output of Au+ beam current at the tandem accelerator used to measure the LHD plasma potential was increased from about 3 μA to 12 μA and it was demonstrated that the average electron density in the plasma could be measured up to $1.75\times10^{19}$m$^{-3}$. This method is effective and widely applicable to improve the performance of low-energy heavy-ion beam transport systems output from the first stage of tandem accelerators and ion sources by adding a lens effect to the multi-stage accelerator tubes.

## I. INTRODUCTION

A wide range of device development and plasma physics research has been conducted to realize fusion power generation. Various instruments have been developed to diagnose the state of high-temperature plasmas in the field of heat and particle transport in high-temperature plasmas. In thermonuclear fusion plasmas, the radial electric field and its fluctuations in the plasma are considered to affect the plasma confinement performance, and the heavy ion beam probe (HIBP) system has been developed as an instrument to measure these physical quantities. Because the Large Helical Device (LHD) confines the plasma in a magnetic field of nearly 3 Tesla, it is necessary to use heavy-mass ion beams with MeV energy to measure the potential at the plasma center. The LHD HIBP system with a tandem accelerator capable of applying a nominal 3 MV terminal voltage to Au+ ions with energies up to 6 MeV is utilized to generate the probe beam. The probe beam injected into the plasma collides with electrons and ions constituting the plasma and ionizes to Au2+. By analyzing the beam energy of Au2+, which is moderated by the plasma potential at the ionization position, the potential in the plasma can be measured[1],[2]. The HIBP system has been used to measure the radial electric field during impurity transport [1], to study EGAMs excited by energetic ions [1], and for heat transport studies [3]. In addition, the detection of zonal flows in CHS [4], investigations of GAM-derived potential fluctuations in T-10 tokamak, and Alfven Eigen mode-derived potential fluctuations in TJ-II have been reported [5].

In order to improve the performance of the LHD-HIBP system, a negative ion source has been developed [6]. Its spatial resolution has been evaluated [7]. The conversion efficiency of Au negative ions to positive ions in a tandem accelerator has been assessed, and the ionization cross-section of the Au probe beam in collision with protons has been derived and evaluated [8]. Using ionization cross sections, it has been found that electron ionization is dominant for heavy ion beams in the keV region, but that proton collisions are more dominant than electron collisions in the MeV region.

This paper reports on the introduction of a new cesium sputter type negative ion source instead of the plasma sputter type negative ion source that has been developed to improve the S/N of the detection signal by increasing the probe beam current of the LHD-HIBP, and on the development study to improve the beam transport efficiency at the low energy side of the tandem accelerator. First, the beam extraction test of the newly introduced Cesium sputtering negative ion source on the test stand was performed, and the results of the Au negative ion beam current obtained are presented. Next, the new negative ion source was installed in the tandem accelerator of the LHD-HIBP. The beam output characteristics are shown, and the beam current that can be injected into the tandem accelerator and the improvement method to approach the beam current characteristics on the test stand are presented. In order to improve the beam transport efficiency at the injection side of the tandem accelerator, the space electrification and lensing effects of the multi-stage accelerator tubes are verified by numerical simulation using IGUN. Based on the results, the beam transport efficiency of the tandem accelerator is improved by adding lensing effect to the multi-stage accelerating electrodes, which improves the input and output beams of the tandem accelerator without modifying the beamline. The density dependence of the detected beam current and the measurement results of the LHD plasma potential distribution are shown at the end.

## II. LOW ENERGY BEAM TRANSPORT IN HIBP SYSTEM

### A. Negative Ion Beam Extraction test on a test stand

The cesium sputtering negative ion source (High Voltage Engineering, Model 860C Multi Sample Source) uses a thermal filament to ionize cesium vapor and sputter targets to generate a negative ion beam. The target is a revolver type, and 10 metal targets can be loaded. This negative ion source has been reported to produce an Au-beam output of 290 μA at Brookhaven National Laboratory [9].

Before integrating the negative ion source into the LHD-HIBP system, the beam extraction characteristics of the negative ion source were investigated on the test stand, so that the distance from the negative ion source extraction electrode to the Faraday cup FC0 on the beam orbit in the LHD-HIBP system was the same as that from the extraction electrode to the Faraday cup FC0. The beam current distribution was measured 590 mm downstream from the extraction electrode in the test stand.

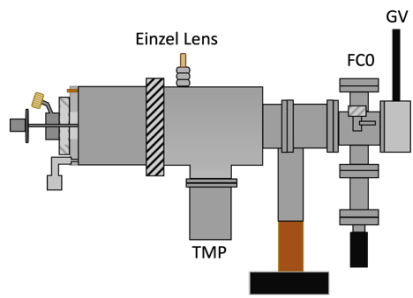

**FIG. 1.** Schematic diagram of the test stand

Fig. 2 shows the characteristics of the negative ion beam current as a function of the extraction voltage for the Model 860C negative ion source. The Einzel lens at the exit of the negative ion source is adjusted to obtain the maximum negative ion beam current in the Faraday cup. Although the beam energy is different from that of a conventional plasma sputtering negative ion source and cannot be precisely compared, it was confirmed that the negative ion beam current was twice as large for Cu negative ions and about the same for Au negative ions.

Based on the results of this test stand, this negative ion source was transferred to the tandem accelerator at LHD-HIBP.

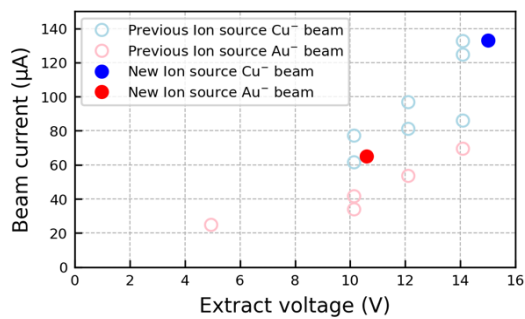

**FIG. 2.** Comparison of beam currents of the old and new Au- and Cu- negative ion sources on the test stand (o: Previous negative ion source [2], ●: new negative ion source).

## B. Tandem accelerator at LHD-HIBP system

The injection beamline to the LHD-HIBP tandem accelerator is shown in Fig. 3(a). Au negative ion beams extracted at 10~20 keV from the negative ion source are focused with an Einzel lens and accelerated to approximately 50 keV in a multi-stage accelerator. Then, only gold or copper negative ions are selectively passed through the sector magnet and injected into the tandem accelerator. In the beamline between the negative ion source and the tandem accelerator, the original performance of the negative ion source obtained on the test stand has not been achieved. In fact, it is necessary to solve the problem that a negative Au ion beam of 100 μA at FC0 at the exit of the negative ion source is reduced to 20 μA at FC1 before the injection of the tandem accelerator due to beam loss in the transport line. Between FC0 and FC1, there is a multi-stage accelerator tube for beam acceleration and a sector magnet for mass separation. Adjustment of the beam focus with an Einzel lens downstream of the negative ion source does not increase the negative ion beam current at FC1. It is not easy to modify the beamline components and sector magnet poles.

In order to improve the beam transmittance of this system, a multi-stage accelerating tube was designed to provide electrostatic acceleration as well as electrostatic lensing. This method can provide both acceleration and focusing effects at the same time when there is no space to add new beam control components to the beamline.

Fig. 3(b) shows a cross-section of a multi-stage accelerating tube. The multi-stage accelerating tube consists of four electrodes, with neighboring electrodes connected by resistors. The voltages of electrodes 1, 2, 3, and 4 are V1, V2, V3, and VG, respectively, starting from the downstream of the negative ion source. A resistor of 335 MΩ is attached between each electrode, and applying a high voltage of 44 to -48 kV between electrode 1 and electrode G results in the equal voltages V1 - V2 = V2 - V3 = V3 - VG. The beam energy when extracted from the negative ion source was set to 16-20 keV, and the negative ion beam energy injected into the tandem accelerator was fixed at 64 keV.

The voltage applied to each electrode of a multi-stage accelerator can be varied to accelerate and converge the negative ion beam. The beam current and divergence angle of the negative ion beam entering the multi-stage accelerator are necessary to obtain the optimum value of the voltage divider by calculating the trajectory of the negative ion beam. To obtain these parameters, negative ion beam extraction experiments were conducted.

a)

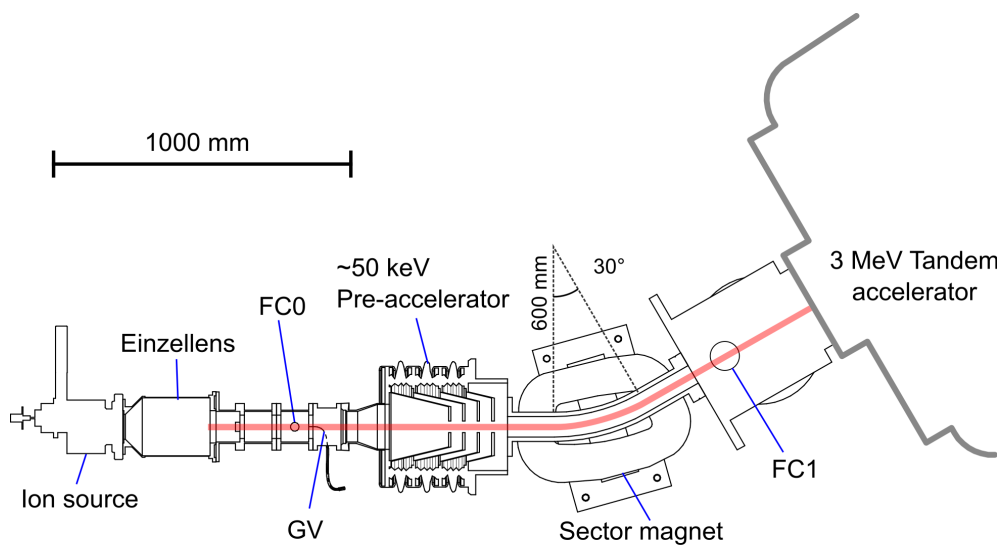

b)

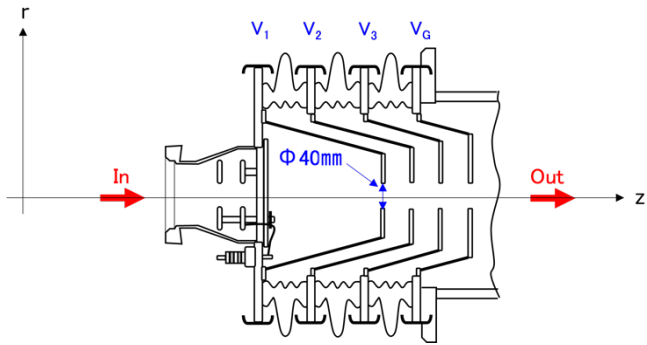

**Fig. 3.** (a) Schematic of the low-energy beamline upstream from the tandem accelerator in the LHD-HIBP system. [8]. FC0: Φ30mm slit for distribution measurement with a linear induction machine, FC1: Φ30mm Faraday cup. (b) Schematic view of the inside of the multi-stage accelerating tube. .

Fig. 4 shows the measured Au- beam current distribution in the Faraday cup FC0 (inlet slit diameter Φ30 mm, beam direction z = 530 mm, z = 0 mm is the target position of the sputter ion source). The current distribution of the negative ion beam was measured by sweeping the Faraday cup up and down. The full width at half maximum of the Au- negative ion beam profile of the new negative ion source was found to be 24 mm, which is almost the same beam diameter as the beam profile of the conventional ion source [2].

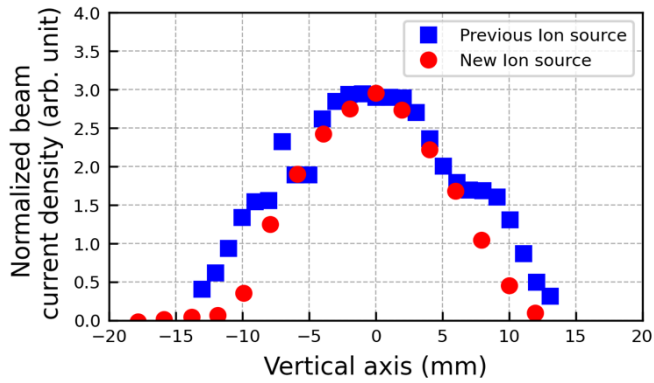

**Fig. 4.** Comparison of Au- beam current distribution at FC0 between the previous and new negative ion sources. The old negative ion source had an Au- beam energy of 13.9 keV. The new negative ion source has an Au- beam energy of 21 keV.

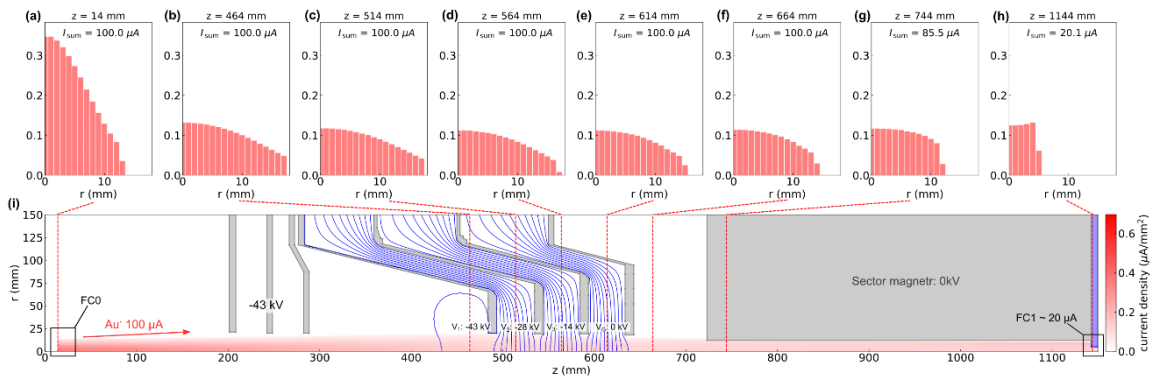

**FIG. 5.** Model and example of ion trajectory calculation used to optimize the voltages V1, V2 and V3 for a multistage accelerator tube, modeled from FC0 to FC1, z = 0 mm: FC0, z = 489 mm: V1 at electrode 1, z = 539 mm: V2 at electrode 2, z = 589 mm: V3 at electrode 3, z = 638 mm VG to electrode 4. The upper panel shows the negative ion beam distribution at each position.

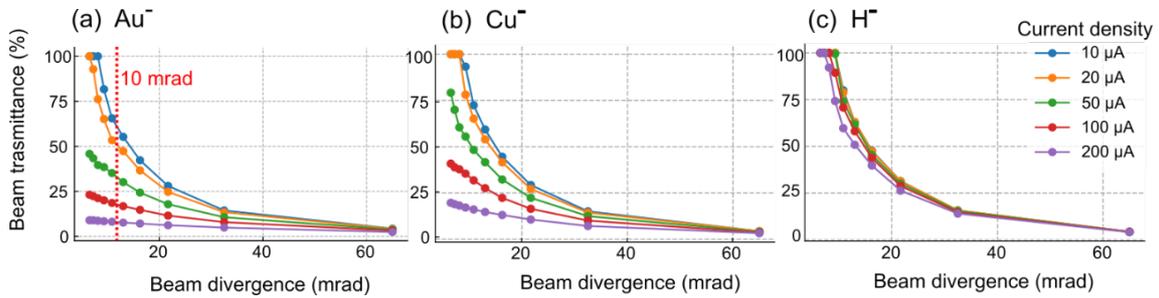

**FIG. 6.** Dependence of negative ion beam transmittance on incident divergence angle and beam current. (a) Au negative ion beam, (b) Cu negative ion beam, and (c) H negative ion beam with an initial negative ion beam energy of 21 keV and an applied voltage of V1=V2=V3=14 kV for multi-stage acceleration.

### C. Low energy beam transport using IGUN

IGUN, an ion source design code, was used to calculate the ion beam trajectory from FC0 to FC1 before the entrance of the tandem accelerator. In this paper, the sign of the voltage parameters used in the calculations are reversed. In a cylindrical coordinate system with the z-axis in the beam direction, the position of the Faraday cup FC0 downstream of the negative ion source extraction electrode is z = 530 mm and the position of the Faraday cup FC1 in front of the tandem accelerator is z = 2250 mm when the target position is z = 0. The three electrodes at the beam orbital position of the multistage electrode have a hole diameter of 40 mm and are located at z=539 mm, 589 mm, and 638 mm. A typical IGUN calculation using a Au ion beam is shown in Fig. 5. Under these conditions, the total acceleration voltage was -43 kV and the voltage between the electrodes was V1- V2 = V2- V3 = V3- VG. The total beam current was set to 100 µA, the same as in the experiment, and the initial beam divergence angle was set to 10 mrad, which corresponds to a negative ion beam current of 20 µA in FC1, obtained experimentally. From the negative ion beam trajectory calculation, it was found that the beam is lost at the electrodes of the LHD-HIBP multi-stage accelerator system and at the Sector Magnet, which performs mass separation. Next, the dependence of the beam transmittance on the initial divergence angle is shown in Fig. 6. For Au-, Cu-, and H-, the transmittance decreases as the divergence angle increases. For Au-, Cu-, and H-, the transmittance decreases as the divergence angle increases, indicating that the space-charge effect decreases as the ion beam current increases for divergence angles below 30 mrad. Under these conditions, the beam transmission is about 20%. From Fig. 6(a), when the beam current is 100 µA, the beam divergence is considered to be caused mainly by the space-charge effect. Therefore, we investigated whether the increase in beam diameter due to the space-charge effect could be suppressed by optimizing the applied voltage of each electrode. Fig. 7 shows the transmittance of the negative Au ion beam when the beam current is 100 µA, the divergence angle is 10 mrad, the voltage $V1 = -43$ kV is fixed, and the voltages $V2$ and $V3$ are swept. From these results, it was found that the combination of equal voltages between each electrode by conventional resistive division exists in the region of low transmittance, and that the transmittance suddenly becomes high at the moment of crossing a certain boundary. Therefore, it was found that there is an optimal combination of voltages that improves the transport efficiency of negative ion beams compared to the conventional condition of equal voltage ratio in a multi-stage accelerating tube.

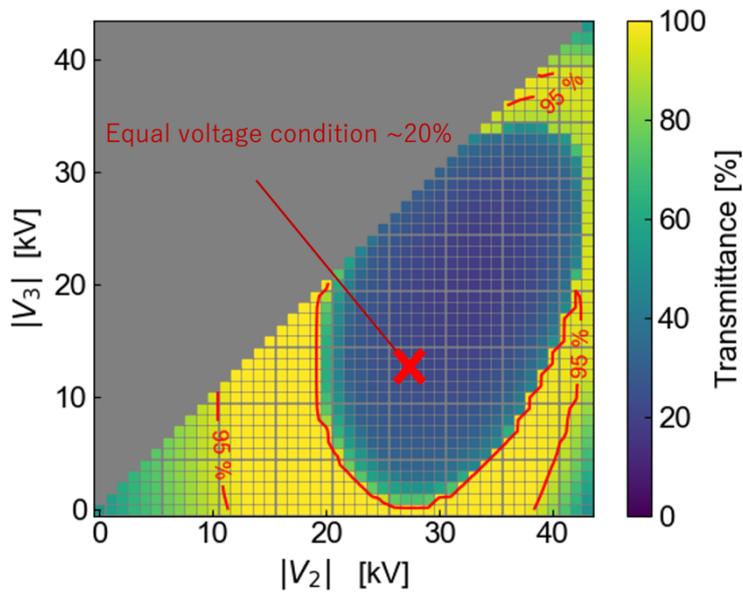

**Fig. 7.** Transmittance maps for electrode voltages V2 and V3 for multi-stage acceleration

## III. Experiments on optimization of negative ion beam transport with multiple acceleration voltages

Based on the results of the previous ion beam orbit calculation, the applied voltage of the multi-stage acceleration electrodes was adjusted to have a lensing effect in order to reduce the transport loss of negative ion beams. The resistors attached to the multi-stage acceleration electrodes were removed, a high-voltage power supply was attached to electrodes 2 and 3, and a search for the optimum multi-stage acceleration voltage was conducted based on the results of the previous numerical negative ion beam orbit calculation. The Au negative ion beam energy injected to the multi-stage electrodes was set to 17 keV, and the potential difference between the electrode voltages V1 and VG was fixed at 47 kV. The resulting transmission coefficients of the negative ion beam for V2 and V3 are shown in Fig. 8. The operating parameters of the negative ion source were fixed at -6 kV and 1.13 mA for the target power supply and 9.81 V and 16.92 A for the ionizer power supply. The extraction voltage was set to -15.7 kV, the Au negative ion beam energy was 21.7 ke, and the voltage of the Einzel lens downstream of the negative ion source was -17.6 kV.

In the case of the Au target, the beam current was largest at V2=28 kV and V3=0.3 kV. The beam current, which was -32.7 µA at the conventional electrode voltage, was -57.46 µA when V2 and V3 were changed, an increase of approximately 1.76 times. The negative ion beam current is likely to increase if the V3 sweep is made more negative.

In the case of the copper target, the beam current value was the highest at V2=33 kV and V3=5 kV, increasing from -15.4 µA to -19.79 µA by a factor of about 1.29, showing a similar trend to that of the Au- beam.

(a)

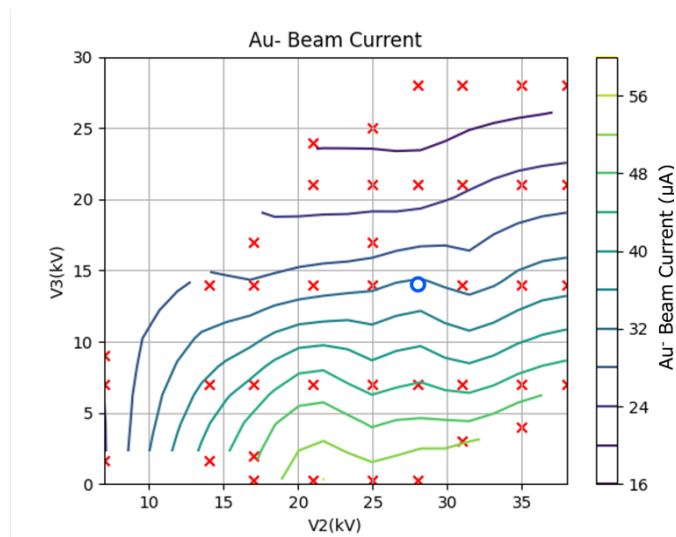

(b)

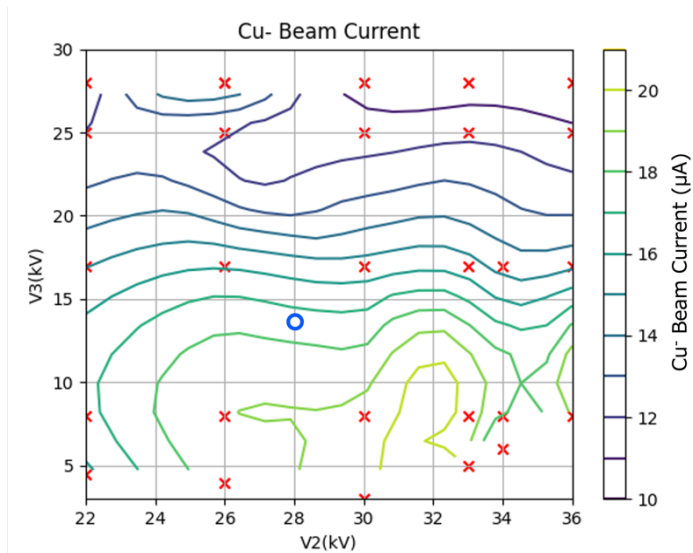

**Fig. 8.** Dependence of negative ion beam current on multi-electrode voltages V2 and V3. The target in (a) is Au, and the target in (b) is Cu. (The unfilled blue circle is equal voltage condition.)

Fig. 9 shows a comparison of the Au negative ion beam current measured at FC1 before and after the optimization of the multi-stage acceleration electrodes in the LHD plasma experiment. Because of the variation in the operating parameters of the Au negative ion source, the target current flowing through the target was used as the physical quantity related to the amount of Au negative ions produced on the abscissa axis. In general, for a constant cesium introduction amount, the higher the target current, the more Au negative ions are produced, and thus the negative ion beam current also increases. This trend was observed even before optimization. In LHD experiments, the negative ion beam current increases by about a factor of two from 38 µA to a maximum of 81 µA in the range of 1 mA to 1.5 mA target current due to the power supply capacity of the beam transport system from the tandem accelerator to the LHD injection. In LHD experiments, the negative ion beam current is limited above 1.5 mA due to the power supply capacity of the beam transport system from the tandem accelerator to the LHD.

Fig. 10 shows the relationship between the Au positive ion beam produced after passing through the 3MV tandem accelerator and the injected Au negative ion beam. The Au+ beam current has increased to 8-12 µA from 30-40 µA for the Au- beam and 3-4 µA for the Au+ beam. The Au- beam current is over 70 µA and the Au+ beam current has decreased to about 8 µA due to the limitation of the beam power due to the power supply capacity as well as the reasons mentioned above.

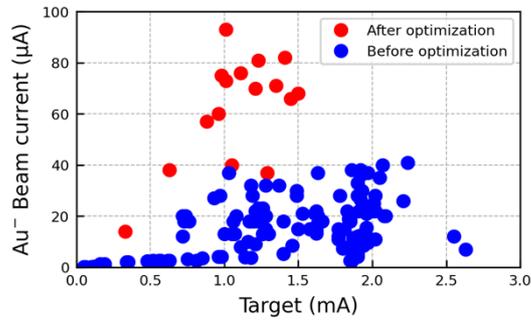

**Fig.9.** Characteristics of Au-beam current of FC1 versus target current of negative ion source before and after optimization of multi-stage acceleration electrodes.

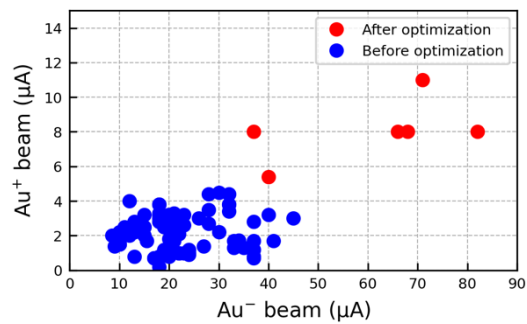

**FIG.10.** Au+ beam current produced after passing through the tandem accelerator versus Au- beam current in FC1

Fig. 11 shows the LHD plasma potential measurement results after improving the negative ion beam transport efficiency of the HIBP low-energy beamline. The horizontal axis is the normalized minor radius, defined as the effective radius divided by the minor radius. The Au+ beam energy injected into the plasma is 4.538 MeV. The spatial distribution of the injection and secondary beam was obtained by sweeping the voltage of a beam steering electrode mounted near the LHD port with a 10 Hz triangular wave. LHD has a magnetic axis $R_{ax}$ = 3.6 m and a magnetic field strength Bt = 2.75 T. Electron cyclotron heating begins at t = 3.0 s. At t = 3.3 s, the neutral beam is superimposed and remains so until t = 6.0 s. The electron heating is dominant. From t = 6.0 s, electron cyclotron heating is turned off and the overall potential becomes negative, with a flat potential distribution at t = 7.0 s. The electron heating is predominant, resulting in a positive plasma potential toward the center and a positive radial electric field. Next, the dependence of the signal intensity $I_{sum}$ of the secondary beam $Au^{2+}$ detected by the energy analyzer on the line-averaged density $\overline{n_e}$ [m$^{-3}$] measured by the interferometer is shown in Fig. 12. The spatial distribution was measured by sweeping the incident beam during the measurement, and is plotted for each region of the plasma normalized minor radius r/a. The signal near $\overline{n_e}$=0 indicates $I_{sum}$ due to collisions with neutral particles before the plasma discharge. The hatched area indicates the noise level $I_{sum}$ <0.25 during the discharge. In this discharge, $I_{sum}$ is smaller for r/a<-0.2 than for the other regions, but there is little difference in $I_{sum}$ for r/a>-0.2. This result is obtained from a typical LHD discharge and argues that the plasma potential can be measured under conditions where the density is up to $\overline{n_e}$=1.75x10$^{19}$m$^{-3}$

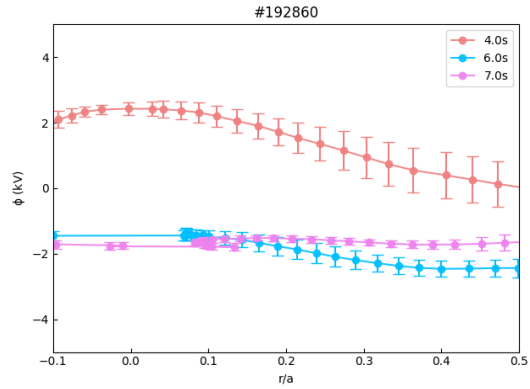

**FIG. 11.** Plasma potential distribution

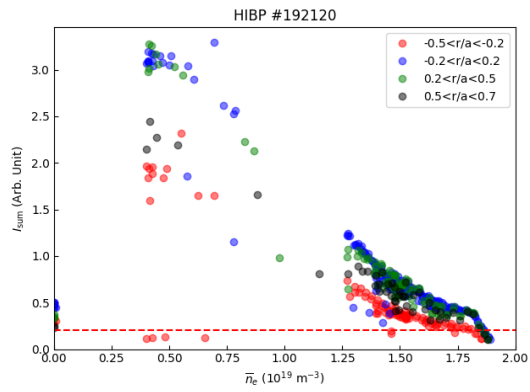

**FIG. 12.** Dependence of secondary beam signal intensity on the electron density

## IV. SUMMARY

In this study, we introduced a new negative ion source and improved the negative ion beam transport system to take advantage of its performance in order to improve the performance of the LHD-HIBP system. The newly introduced cesium sputter type negative ion source was tested on a test stand, and it was confirmed that a gold negative ion beam current of more than 100 μA could be obtained. By introducing this negative ion source into the LHD-HIBP system and adding a lensing effect to the multi-stage accelerator electrodes, the negative ion beam current injected into the tandem accelerator was increased by a factor of 3.6. As a result, the Au+ beam current could be used for HIBP plasma potential measurement by a factor of 2-3. The method of improving the beam transport efficiency by the electrostatic lensing effect in a multi-stage accelerator is useful in many accelerator systems because it does not require any reconfiguration of the beam transport system and allows an accelerator design with fewer components.

## V. ACKNOWLEDGMENTS

This work was supported by the JSPS KAKENHI under Grant Nos. 19KK0073 and 23H01160. We would like to thank the LHD experiment group for their generous experimental support and fruitful discussions, and Mr. Y. Nakajima for his encouragement. One of authors (K.N.) is grateful to Dr. M. Okamura at Brookhaven National Laboratory for his valuable discussion.